# Pressure-induced electro-switching of polymer/nano-graphene composites

Eirini Kolonelou [1], Anthony N. Papathanassiou [1, *] and Elias Sakellis [1,2]

[1] National and Kapodistrian University of Athens, Physics Department. Panepistimioupolis, GR 15784 Zografos, Athens, Greece

[2] National Center of Natural Sciences Demokritos, Institute of Nanomaterials and Nanotechology, Aghia Paraskevi, Athens, Greece

**Abstract:**

The pressure and temperature dependency of the electrical conductivity of poly(vinyl alcohol)/poly(vinyl pyrrolidone) (1/1, w/w) and poly(vinyl alcohol) and composites with dispersed nano-graphene platelets were studied. Above the critical platelet fraction for electric charge percolation, the composites function as pressure-induced electro-switches. The conductor to insulator transition is optimally intense and stable. The electrical conductivity drops by two orders of magnitude at a critical pressure around 750 bars. The transition is stable over tenths of degrees above room temperature. The reduction of the conductivity upon pressure results from the competition between the pressure dependencies of the polarizability of the polymer matrix and the inter-platelet separation, respectively. Both contributions control the fluctuation induced tunneling of electrons through the polymer barrier separating adjusting conductive platelets. The role of the local electric field at the polymer-platelet interfaces by assisting tunneling is suppressed by the decrease of the polarizability upon pressure.

# I. Introduction

Flexible polymer composites with dispersed metal, inorganic or carbon allotropes nanoparticles have been used to develop pressure sensors and electro-mechanical switches. Depending on the polymer used, these structures are employed for different in different applications, such as micro-electronic and biocompatible human monitoring devices [1, 2, 3]. . In general, switching properties result from the shortening of the distance between neighboring inclusions. Recently, novel polymer-based switches and sensors based on the tuning of the electron tunneling through the polymer spacing were developed [4]; pressure yields an increased conductivity state. In the present work, we address the possibility of a polymer based composite which operates in an *inverse* manner on compression; i.e., switching from a conducting to an insulating state on pressurization. To achieve it, we dispersed nano-graphene platelets (NGPs) in a polar polymer, whereas its polarizability decreases on pressure. Consequently, the current in the tunneling region is suppressed on compression.

Poly(vinyl alcohol) (PVA) is a water-soluble synthetic polymer, which has excellent film forming, emulsifying and adhesive properties. It has high tensile strength and flexibility comparable to those of human tissues. PVA finds its wide application in plasterwork and joint sealing due to its favorable properties of weather resistance, waterproof, non-swell with water, non-embrittlement, non-poison, tastelessness, and low cost. [5]. Polyvinylpyrrolidone (PVP) is water-soluble optically transparent polymer. [6] PVP is soluble in water and other polar solvents, such as alcohols. It has excellent wetting properties and readily forms films. This makes it good as a coating or an additive to coatings. [7]. Graphene is an allotrope of carbon consisting of a single layer of carbon atoms arranged in an hexagonal lattice. [8]. It is the strongest material ever tested, efficiently conducts heat and electricity and is nearly transparent. Graphene nano-ribbons, graphene nano-platelets, and nano–onions are believed to be non-toxic at concentrations up to 50μg/ml [9].

The glass transition temperatures of neat PVA and PVP are: $T_{g,PVA}$=358 K and, $T_{g,PVP}$=441 K, respectively. The glass transition temperature of (PVA/PVP) (1/1, w/w) can be evaluated through the Fox formula [10]: $T_g^{-1} = w_{PVA} T_{g,PVA}^{-1} + w_{PVP} T_{g,PVP}^{-1}$, where $w_{PVA}$, $w_{PVA}$ denote the mass fractions of PVA and PVP, respectively [11]. For $w_{PVA}=w_{PVP}=0.5$, the glass transition temperature of the blend is estimated: $T_g$=386 K. Nano-graphene platelets (NGPs) constitute the conductive inclusions dispersed into (PVA/PVP) (1/1, w/w). The electric charge flow in (PVA/PVP) (1/1, w/w) and PVA composites loaded with different mass fractions of NGPs is explored at various spatiotemporal scales and isothermal and isobaric conditions below $T_g$, by employing Broadband Dielectric Spectroscopy (BDS). Polymer composites loaded with carbon allotropes exhibit low critical mass fraction values for percolation of electrical current [12 - 17]. Dc conductivity is governed by fluctuation induced tunneling (FIT) of electrons through the polymer separating neighboring carbon allotrope grains and through the potential barrier settled by physical contact of some inclusions. Polarization of the polymer within the tunneling region enhances conduction [15].

In the present work, we report an electro-switching property of the composites induced by hydrostatic pressure, which is optimal for NGP loading just above the critical percolation concentration. Pressure drives the composite from an (electrically) conducting to an insulator phase, which contradicts the common sense that, around the critical percolation point, compression brings the conducting island closer and thus an onset of electrical conductivity should occur. The phenomenon is interpreted by the reduction of the polymer polarization and tunneling assisting local electric field intensity at the NGP-polymer interfaces upon pressure.

## II. Experimental details

Equal masses of PVA and PVP powders were dissolved separately into warm double distilled water by continuous stirring. The aqueous solutions merged together. NGP powder was dissolved into water at 353K and ultra-sonicated in a heat bath for about two hours. Subsequently, the aqueous solutions were mixed, and stirred, prior to drop-casting on a teflon surface. After 48 hours drying at ambient conditions, free standing specimens of about 1 mm thickness were obtained. SEM microscopy ensured a homogeneous distribution of the NGPs [12]. The same procedure was employed to obtain PVA matrix. A Novocontrol High Pressure BDS system was used for dielectric measurements at temperatures above the ambient one and pressure less than 3 kbars, in the frequency range from $10^{-2}$ to $10^6$ Hz. Samples of typical surface area of 1 cm$^2$ and thickness of 1 mm were placed in the pressure vessel following the methodology published earlier [18 - 20]. Complex permittivity measurements were collected from 1 mHz to 1 MHz with a Solartron SI 1260 Gain-Phase Frequency Response Analyzer, equipped with a Broadband Dielectric Converter (BDC, Novocontrol). Data acquisition was monitored through the WinDeta (Novocontrol) software [21, 22]. The formalism of the complex permittivity ε* and the electric modulus M* as a function of frequency f were employed to study the effect of temperature, pressure and composition on the dielectric properties of the composites.

## III. Experimental results

The imaginary part of the complex permittivity function $\varepsilon^*(f) = \varepsilon'(f) + i\varepsilon''(f)$, where $i^2 = -1$ and f is the frequency of an externally applied harm, was employed:

$$\varepsilon''(f) = \frac{\sigma_{dc}}{\varepsilon_0 2\pi f^n} + \frac{\Delta\varepsilon}{\left(1+(f/f_0)^a\right)^b} \tag{1}$$

where $\sigma_{dc}$ denotes the dc-conductivity, $n$ is a fractional exponent ($n \leq 1$) which is usually close to 1, $\Delta\varepsilon$ is the intensity of a relaxation mechanism, $a$ and $b$ are fractional exponents and $f_0$ is a parameter that coincides with the peak maximum frequency when $b=1$. In the present work, the full spectra are fitted by eq. (1) and we focus on the values of $\sigma_{dc}$. The dielectric spectra of poly(vinyl alcohol)/poly(vinyl pyrrolidone) (1/1, w/w) with different loadings of nano-graphene platelets; i.e., x= 0.05, 0.1 and 0.3 w/w % NGP, recorded at T=293 K and different pressure are depicted in Figure 1. We observe that, for given P-T conditions, the dc-conductivity, appearing as a low-frequency, straight line, is enhanced by the increase of NGP content. The threshold for electric charge percolation is determined around 01 w/w % NGP [23]. The rate at which the plots decrease upon compression is not constant, revealing their distinct grouping into a couple of sets, for x= 0.3 w/w % NGP. The separation is about two-orders of magnitude. To ensure that the phenomena observed are bulk ones, we alternatively used the complex electric modulus function: $M^* = \frac{1}{\varepsilon^*} = M' + iM''$ [24], where:

$$M'(f) = M_s \frac{(f\tau_\sigma)^2}{1+(f\tau_\sigma)^2} \tag{2}$$

$$M''(f) = M_s \frac{f\tau_\sigma}{1+(f\tau_\sigma)^2} \tag{3}$$

A plot of $M''(f)$ against $logf$ consists of a low frequency "conductivity peak", which stems from the $\sigma_{dc}$ component itself, with a maximum at $f_{0,\sigma} = (2\pi\tau_\sigma)^{-1}$, where $f_{0,\sigma} = 2\pi\sigma_{dc}/(2\pi\varepsilon_\infty)$ $f_{0,\sigma} = \sigma_{dc}/(2\pi\varepsilon_\infty)$ and $\varepsilon_s \equiv \varepsilon'(f \to \infty)$. The high frequency plateau yields $M_s \equiv M'(f \to \infty)$ and the (relative) static dielectric constant $\varepsilon_s = \frac{1}{M_s}$, as well. While complex permittivity measurements may suffer from undesirable low frequency space charge capacitance contributions, the complex electric modulus formalism suppressed them and permits a clear determination of both dc conductivity and static dielectric constant $\varepsilon_s$. As can be seen in Figure 1, the grouping of the $\varepsilon''(f)$ curves is alternatively confirmed in the $M''(f)$ representation, as the 'conductivity peak' distribute in two district sets.

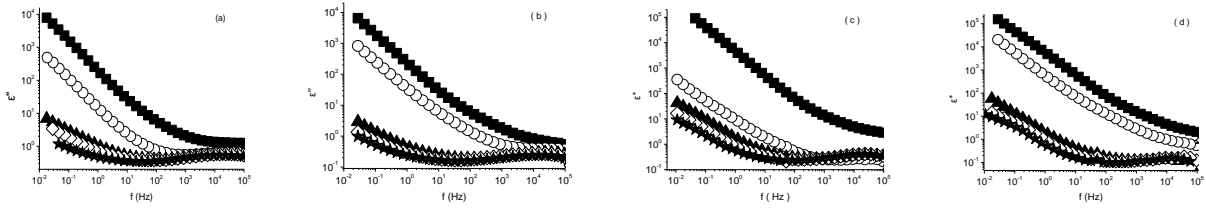

*Figure 1. The dielectric spectra of (PVA/PVP) (1/1, w/w): NGP composites: (a) 0 w/w %, (b) 0.05 w/w %, (c) 0.1 w/w %, (d) 0.3 w/w % NGP, recorded at T=313 K and different pressure states: 1 bar; circles: 500 bar; triangles: 1000 bar; diamonds: 1500 bar; stars:2000 bar.*

For temperatures below 373 K, different isobars have the tendency to group in two sets (Figure 3). At temperatures close to $T_g$, the initiation of co-operative macromolecular mobility affects the critical behavior of the percolation network formed above, and near the percolation threshold. In Figure 4, the logarithm of the dc conductivity, extracted from the analysis of the $\varepsilon''(f)$ spectra, is depicted against pressure for 293K and 313 K, for the entire set of (PVA/PVP) (1/1, w/w: NGP composites. For pure (PVA/PVP) (1/1, w/w) and composites with NGP loading less than 0.3 w/w %, the data points distribute smoothly over a band spanning over an order of magnitude. For 0.3 w/w % NGP mass fraction, $log\sigma_0(P)$ can be approximated by a step-function centered at about $P_c \approx 750\,bar$ and height about 2. Hence, the phase transition induces a two orders of magnitude reduction of the electrical conductivity. Although the percolation limit was determined around 0.1 % w/w NGP, the pressure induced electro-switching feature is pronounced for x=0.3 w/w % NGP composites, at a critical pressure $P_c$. At this NGP loading the functionality was practically stable at temperatures up to about 60 degrees above room temperature.

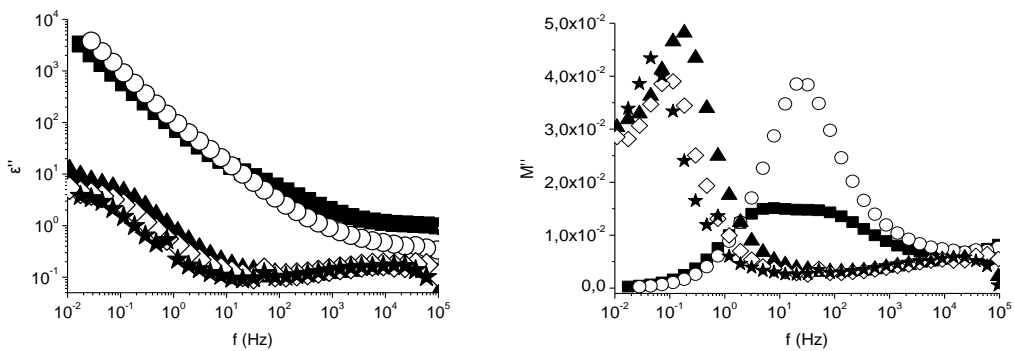

*Figure 2. Comparative presentation of $\varepsilon''(f)$ and $M''(f)$ of (PVA/PVP) (1/1, w/w) : x=0.3 w/w % NGP composites, recorded at T=293 K and various pressures, to the experimental data points; (squares: 1 bar; circles: 500 bar; triangles: 1000 bar; diamonds: 1500 bar; stars:2000 bar). In the electric modulus representation, the dc-conductivity contribution appears as a "conductivity relaxation peak". Regardless the representation used, accumulation in two groups, is observed.*

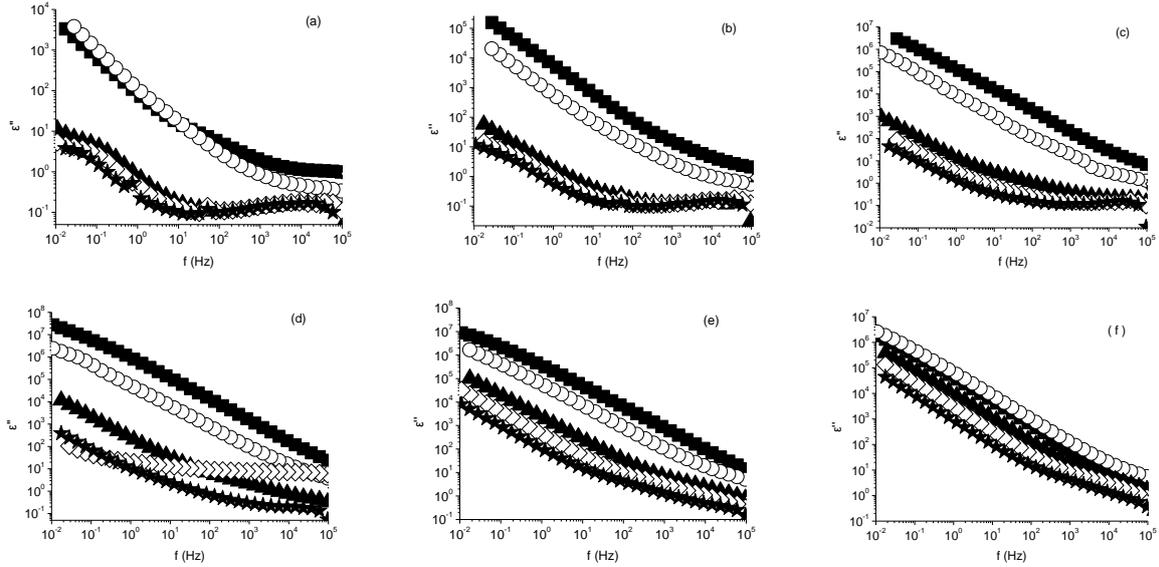

***Figure 3.*** *Effect of temperature on the grouping of isobaric ε''(f) spectra observed in (PVA/PVP) (1/1, w/w): x=0.3 w/w % NGP composites. (a) T=293 K, (b) T=313 K, (c) T=333 K, (d) T=353 K, (e) T=373 K, (f) T=393 K and different pressures. (squares: 1 bar; circles: 500 bar; triangles: 1000 bar; diamonds: 1500 bar; stars:2000 bar). At temperature close to room temperature, compression above 500 bars, induces a transition from a conductive phase to an insulating one characterized by a step of about two orders of magnitude in the dc conductivity values.*

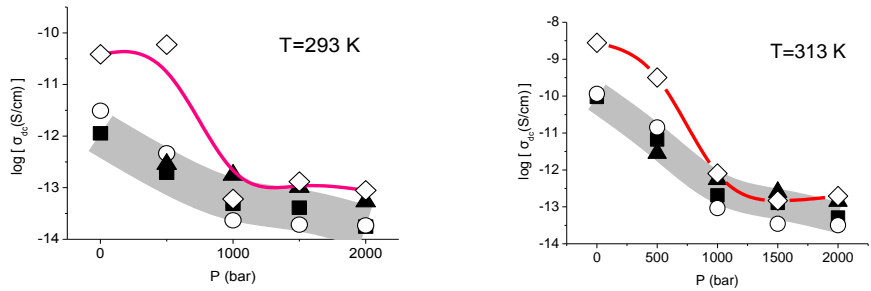

***Figure 4.*** *The logarithm of the dc-conductivity against pressure at different mass fractions of NGP: 0 w/w % (squares), 0.05 w/w % (circles), 0.1 w/w % (triangles), 0.3 w/w % (diamonds), at T=293 K and T=313 K. For x=0.3w/w % NGP, compression yields a step-like transition of the conductivity.*

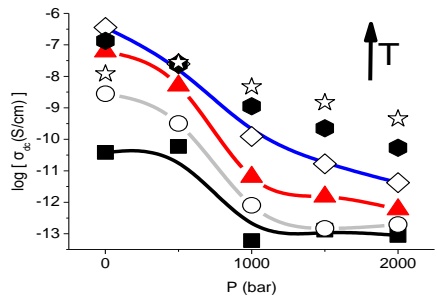

***Figure 5.*** *The logarithm of the dc-conductivity against pressure at different temperatures: T=293 K (squares), T=313 K (circles), T=333 K (triangles), T=353 K (diamonds), T=373 K (hexagons), T=393 K (stars); for (PVA/PVP) (1/1, w/w) : x=0.3 w/w % NGP composites. Note that the pressure induced step-like behavior is pronounced on approaching room temperature.*

For x=0.3 w/w % NGP, the switching functionality is sufficiently stable up to *373K* (Figure 5). The composite loses its functionality at temperatures, close to the vicinity of $T_{g,PVA-PVP}$, can likely be explained by the initiation of co-operative motion of macromolecules, which perturbs the pressure – induced criticality. Moreover, the instability at temperatures exceeding 60 degrees above room temperature, may also be correlated with the local softening of PVA domains while the blend remains globally glassy [23].

PVA (x w/w % NGP) composites exhibit electro-switching behavior similar to that of PVA/PVP: NGP composites. Similarly, as can be seen in Figure 6, PVA (0.3 w/w % NGP) retains its electro-switching characteristics up to the glass transition temperature of PVA $T_{g, PVA} = 386\ K$.

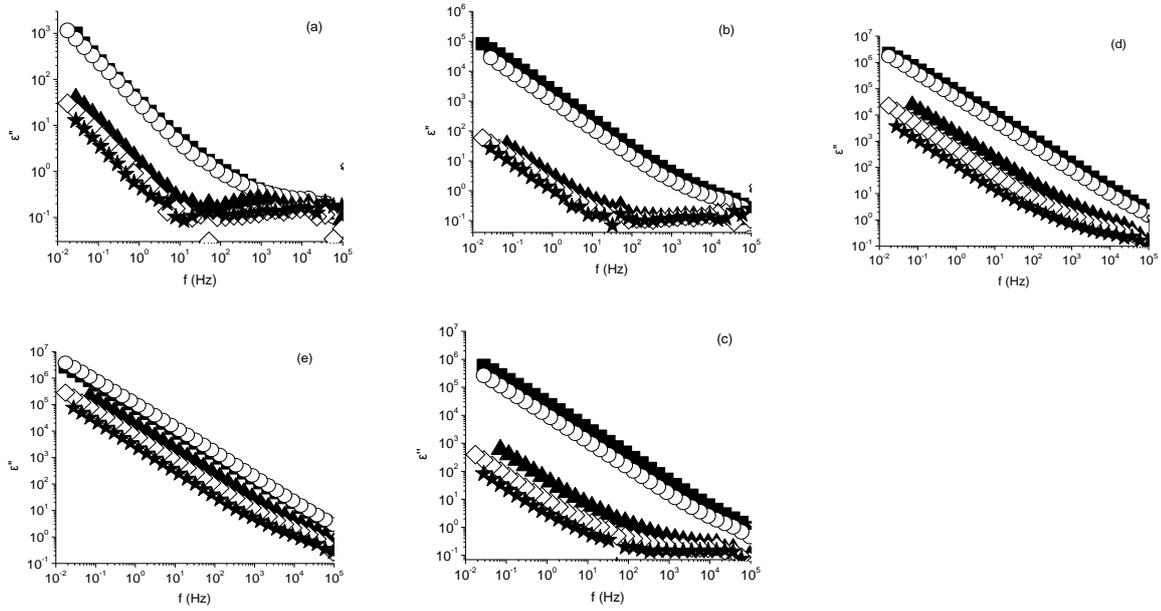

*Figure 6. Dielectric spectra (ε´´) of PVA: 0.3 % w/w NGP recorded at different temperatures [(a) T=293 K, (b) T=313 K, (c) T=333 K, (d) T=353 K, (e) T=373 K] and different pressures. (squares: 1 bar; circles: 500 bar; triangles: 1000 bar; diamonds: 1500 bar; stars:2000 bar).*

## IV. Discussion

Electric charge transport along polymer composites with low fraction of dispersed conducting inclusions occurs by fluctuation induced tunneling (FIT) [25]. According to this model, phonon-induced fluctuations of the Fermi energy of extended electronic states within the conducting islands, assists quantum mechanical tunneling of electrons through the insulating polymer barrier separating neighboring conducting inclusions. Even at NGP fractions higher than the percolation threshold, whereas physical contact of NGPs may occur, the electron transport trough the contact surface of NGP is also governed by FIT [15]. The electrical conductivity is given by:

$$\sigma = \sigma_0 \exp\left(-\frac{T_1}{T+T_0}\right) \quad (4)$$

where $\sigma_0$ is a pre-exponential factor,

$$T_1 = AE_0^2 \varepsilon_0 \varepsilon / k \quad (5)$$

where $A$ denotes the volume of tunneling contact $E_0$ is the mean electric field intensity at the time interval that tunneling occurs, $\varepsilon$ is the relative dielectric constant of the material penetrated, $\varepsilon_0$ is the permittivity of vacuum and k is the Boltzmann's constant and

$$T_0 = T_1 / 2\chi w \xi \quad (6)$$

where $\chi \equiv \sqrt{2mV_0/h^2}$, where $m$ is the mass of the transferring charge, $V_0$ is the height of the inter-NGP potential barrier, and $h$ is the Planck's constant, $w$ is the width of the tunneling path and $\xi$ is a parameter related to the shape of the potential barrier.

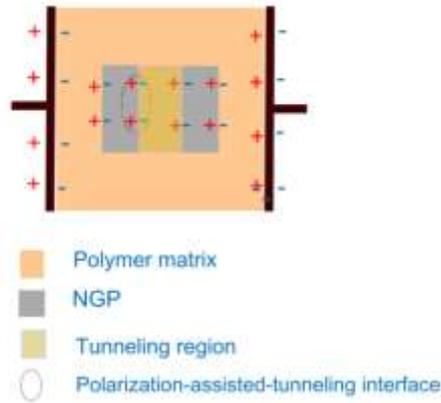

*Figure 7. Simplified sketch cross section of a polymer: NGP composite placed inside the capacitor-type sample holder. Due to the interfacial polarization, an electric field perpendicular to the interface surface area and oriented towards the direction of the applied electric field. Subsequently, tunneling is driven by increased total electric field intensity.*

FIT was initially proposed to model electrical conductivity in granular metals and, later, in electron conducting conjugated polymers with inhomogeneous structural and electrical disorder [26]. The polarization of the polymer matrix of carbon composites [15] induces a local electric field at the conducting inclusions – polymer interface, in the direction of that externally applied. The later enforces FIT through the potential barrier separating neighboring inclusions. An externally applied electric field leads to accumulation of surface charge on either side of the polymer – NGP interface that is non parallel to the vector of the external field. The charge separation between opposing interfaces of neighboring NGPs, dictates an additional electric force along the direction of the external electric field. Therefore, electrons are enforced furthermore to tunnel from one NGP towards a neighboring one.

In principle, the polarizability is a function of temperature and pressure. Isothermal compression of the composites affects fluctuation induced tunneling mechanism mainly in two ways:

**(i):** The polymer matrix is more compressible than NGPs. Therefore, pressure reduces the inter-NGP separation and, the tunneling width shortens. The isothermal compressibility of the polymer can be written approximately as: $\kappa_T \equiv -\left(\frac{\partial \ln V}{\partial P}\right)_T \cong -3\left(\frac{\partial \ln w}{\partial P}\right)_T$, where $V$ denotes the volume of the specimen. Hence, the percentage reduction of $w$ upon a pressure change $\Delta P$, is roughly equal to $3\kappa_T$. In turn, a shortening of $w$, enhances the tunneling current, as a result of stronger overlap of electron wave functions spanning over neighboring NGPs. Using a typical value for the isothermal compressibility $\kappa_T \approx 3 \times 10^{-3} MPa^{-1}$ [27], the percentage variation of the tunneling width can be estimated: $(\Delta w/w)/\Delta P \approx -0.01\% bar^{-1}$.

**(ii):** Compression weakens the role of polymer polarizability on FIT current, due to the reduction of rotational motion of macromolecules. In Figure 7, where a snapshot of a composite under the influence of an external field, bound electric charge layers are formed at the polymer-NGP interfaces; a local electric field is developed along the axis connecting neighboring NGPs, with primary direction identical to that of the external field. Thus, polarization of the polymer volume confined between neighboring NGPs enforces the total driving electric force for electron tunneling. Pressure weakens the role of polymer polarization on the tunneling current. Our results seem to comply with the latter scenario: the static dielectric constant $\varepsilon_s(P)$ (estimated from the low frequency data points of $\varepsilon'(f)$ at various isobars) exhibits negative monotony for both neat PVA and (PVA/PVP) (1/1, w/w) (Figure 8). At $T=313\ K$ and $P \rightarrow 1 bar, we\ get:$ $\left(\frac{\partial \varepsilon_s}{\partial P}\right)_T = -(7.5 \pm 0.1) \times 10^{-3} bar^{-1}$, or $(\Delta \varepsilon_s/\varepsilon_s)/\Delta P \approx -0.08\% bar^{-1}$. Note that the latter values are estimated in the ambient pressure limit, so as to compare with isothermal compressibility values $\kappa_T$ reported for the same pressure limit. We conclude that the term of the percentage variation of the polarization of the polymer spacing is about eight times larger than the percentage variation of the inter-NGP separation $w$, per bar.

When the mass fraction of NGPs is high enough (i.e., beyond the percolation threshold), physical conduct among a fraction of NGPs occurs. FIT through the NGP-NGP junction proceeds, in parallel with tunneling through the polymer. However, at fractions just above the percolation threshold, the inter-NGP links are vary in their quality of mechanical contacts, orientation and geometry of the inter-NGP tunneling region and, therefore, the pressure affects tunneling through the NGP-NGP contacts.

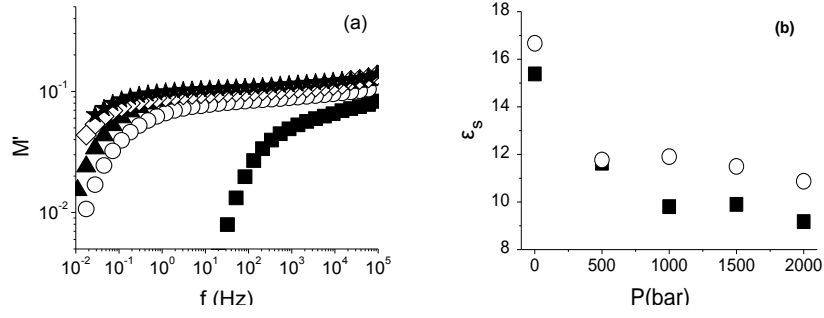

*Figure 8* (a): The real part of the electric modulus M´(f) of neat PVA at 313K and different pressures (squares: 1 bar; circles: 500 bar; triangles: 1000 bar; diamonds: 1500 bar; stars:2000 bar), (b):The high frequency plateau yields $M_s \equiv M'(f \to \infty)$ and, subsequently, $\varepsilon_s = 1/M_s$. Pressure dependence of the relative dielectric constant $\varepsilon_s$ at 313 K for neat? PVA (squares) and PVA-PVP 1-1 (circles).

The charge density in the tunneling volume of the inter-NGP polymer spacer can is [12]:

$$\rho = \varepsilon_0 \varepsilon_s E/w \tag{7}$$

where $\varepsilon_s$ is the relative static dielectric constant of the neat polymer and *E* is the effective driving electric field intensity. Differentiating eq. (7) with respect to pressure, we get:

$$\frac{dlog\rho}{dP} = \frac{dlog\varepsilon_s}{dP} + \frac{dlogE}{dP} - \frac{dlogw}{dP} \tag{8}$$

Assuming that the term $dlogE/dP$ is practically small compared to the other pressure derivatives appearing in the right side of eq. (8), we get:

$$\frac{(\Delta\rho/\rho)}{\Delta P} \approx \frac{(\Delta\varepsilon_s/\varepsilon_s)}{\Delta P} - \frac{(\Delta w/w)}{\Delta P} \tag{9}$$

The percentage variation of the charge density in the tunneling region upon pressure has two competing components: $(\Delta\varepsilon_s/\varepsilon_s)/\Delta P \approx -0.08\%\ bar^{-1}$ and $(\Delta w/w)/\Delta P \approx -0.01\% bar^{-1}$, in the low pressure limit. The assistance of the polymer polarization to FIT process is suppressed upon compression and cannot be compensated by the shortening of the tunneling width *w*, for weight fractions up to about the percolation threshold; indeed, in Figures 6 and 7, the dc conductivity monotonically decreases upon pressure. At mass fraction *x=0.3 %*, the system is close to the critical behavior, and the dc conductivity is still controlled by the above mentioned parameters. Tunneling through the NGP-NGP junction cannot gain the polymer polarization enforcement to tunneling through the polymer separation. The pressure induced electro-switching function is at x=0.3 w/w %, stems probably from the interplay between competing pressure dependencies of tunneling through the polymer barrier and the pressure dependence of direct physical contacting of NGPs.

## V. Conclusions

Poly(vinyl alcohol)/poly(vinyl pyrrolidone) (1/1, w/w) and poly(vinyl alcohol) nano-graphene platelets composites were studied at different pressure - temperature conditions and loadings. The dispersion of NGPs yields a pressure - induced electro-switching property, which is optimally intense and stable for x=0.3 w/w % NGP, i.e., beyond and lose enough to the percolation threshold. On hydrostatic compression, at a critical pressure about 750 bars, the system undergoes a transition from a semi-conducting phase to a two orders of magnitude less conductive one. The functionality is stable at temperatures up to 60 degrees above room temperature. Dc conductivity in polymer based NGP composites occurs by fluctuation induced tunneling (FIT) through the potential barrier set by the polymer separating neighboring NGPs and, at high NGP fractions, through the potential barrier due to the physical contacting among NGPs. Pressure shortens the tunneling width favoring tunneling. FIT is enhanced by the polarization of the polymer volume, through which proceeds. Pressure weakens the tunneling enhancement assisted due to a local electric field developed at the NGP – polymers inter faces. Based on our results, it seems that the weakening of the polarizability factor on compression is mainly responsible for the appearance of the switching behavior. is probably the result of interplay between different modes of tunneling, manifested significantly by the pressure decrease of the polymer polarization. The composites are possible candidates for bio-compatible sensing and switching devices for tissues, organs and bones, due to their flexibility, non-toxicity, low cost and simple preparation and application.

**Acknowledgments**
We are grateful to Th. Areti and E. Roumelioti for sample preparation and acquisition of experimental data.

**References**

[1] Yin He, Yue Ming, Wei Li , Yafang Li, Maoqi Wu, Jinzhong Song, Xiaojiu Li and Hao Liu, Highly Stable and Flexible Pressure Sensors with Modified Multi-Walled Carbon Nanotube/PolymerComposites for Human Monitoring *Sensors* **2018** *18* 1338-1345

[2] Harper Meng, Guoqiang Li, [ review of stimuli-responsive shape memory polymer composites *Polymer* **2010,** *54,* 2199-2221

[3] [Polymer-Nanoparticle Composites: From Synthesis to ModernApplications Thomas Hanemann 1,2, * and Dorothée Vinga Szabó *Materials* **2010,** *3,* 3468-3517

[4] C. Connolly, (2004),"Switches and pressure sensors benefit from novel composite material", *Sensor Review*, **2004,** *24,* 261 - 264

[5] Hallensleben, M., R. Fuss, F. Mummy Polyvinyl Compounds, *Ullmann's Encyclopedia of Industrial Chemistry.* Wiley **2015**, 1-65.

[6] Haaf, F.; Sanner, A.; Straub, F. Polymers of N-Vinylpyrrolidone: Synthesis, Characterization and Uses *Polym. J.* **1985**, *17*, 143-152.

[7] Guoshan, S.; Yannan, L.; Zhongcheng, Z.; Heying, Z.; Jinping, Q.; Changcheng, H.; Huiliang, W. Strong Fluorescence of Poly(N-vinylpyrrolidone) and Its Oxidized Hydrolyzate *Macromolecules (Rapid Communications)* **2015**, *36*, 278-285.

[8] ISRN Condensed Matter Physics *International Scholarly Research Network* **2012**, 1–56.

[9] Talukdar, Y.; Rashkow, J. T.; Lalwani, G.; Kanakia, S.; Sitharaman, B. The effects of graphene nanostructures on HYPERLINK mesenchymal stem cells *Biomaterials* **2014**, *35*, 4863–4877.

[10] Fox, T.G., The composition dependence of glass transition properties Bull. Am. Phys. Soc. **1956**, *1*, 12311357-2

[11] M. N. Cassu, M. I. Felisberti, Poly(vinyl alcohol) and poly(vinyl pyrrolidone) blends: Miscibility, microheterogeneity and free volume change, Polymer **1997**, *8*, 3907-3911).

[12] Lorenz, B.; Orzall, I.; Heuer, H.; J. *Phys. A: Math. Gen.* **1993,** *26*, 4711.

[13] Consiglio, R.; Baker, D.; Paul, G.; Stanley, H.; *Physica A.* **2003**, *319*, 49.

[14] Trachenko, K.; Dove, M.; Geisler, T.; Todorov, I.; Smith, B. *Phys.: Condens. Matter.* **2004***, 16* S2623–S2627.

[15] Syurik, J.; Ageev, O.; Cherednichenko, D.; Konoplev, B.; Alexeev, A. Non-linear conductivity dependence on temperature in graphene-based polymer nanocomposite*, Carbon*, **2013**, *63*, 317-323.

[16] Islam, R.; Papathanassiou, A.; Roch Chan Yu King, Brun, J.; Roussel, F. Evidence of interfacial charge trapping mechanism in polyaniline/reduced graphene oxide nanocomposites, *Appl. Phys. Lett.* **2015**, *107*, 053102.

[17] Lampadaris Ch., Sakellis E., Papathanassiou A. N. Dynamics of electric charge transport in polyvinylpyrrolidone / nano-graphene platelets composites at various compositions around the concentration insulator-to-conductor transition *Appl. Phys. Lett.* **2017**, *110*, 222901-4

[18] Papathanassiou, A.; Sakellis, I.; Grammatikakis, J. Dielectric property of Granodiorite partially saturated with water and its correlation to the detection of seismic electric signals Tectonophysics, **2009**, *512*, 148-152.

[19] Papathanassiou, A.; Sakellis, I.; Grammatikakis, J. Negative activation volume for dielectric relaxation in hydrated rocks *Tectonophysics* **2010**, *490,* 307.

[20] Papathanassiou, A.; Sakellis, I.; Grammatikakis, J. Measurements of the dielectric properties of limestone under pressure and their importance for seismic electric signals Origin, *J. Appl. Geophysics* **2014** *102*, 77-80

[21] Papathanassiou, A.; Plonska-Brzezinska, M.; Mykhailiv, O.; Echegoyen, L.; Sakellis, I. Combined high permittivity and high electrical conductivity of carbon nano-onion/polyaniline composites *Synth. Met.* **2015**, *209*, 583-587

[22] Papathanassiou, A.; Mykhailiv, O.; Echegoyen, L.; Sakellis, I.; Plonska-Brzezinska, M. Electric properties of carbon nano-onion/polyaniline composites: a combined electric modulus and ac conductivity study, *J. Phys. D: Appl. Phys.* **2016** *48*, 285305.

[23] Kolonelou, E.; Papathanassiou, A.; Sakellis, E. Evidence of local softening in glassy poly(vinyl alcohol)/poly(vinyl pyrrolidone) (1/1, w/w) nano-graphene platelets composites, *Mat. Chem. Phys.* (submitted); arXiv:1809.06301v1 [cond-mat.mtrl-sci]

[24] Kremer, F.; Schönhals, A. Broadband Dielectric Spectroscopy *Springer* **1991**, p. 59.

[25] Sheng P. Fluctuation-induced tunnelling conduction in disordered materials. *Phys Rev B,* **1980**, *2* 2180–95.

[26] Kaiser, A.; Skakalova, V. Electronic conduction in polymers, carbon nanotubes and grapheNe, *Chem. Soc. Rev.,* **2011**, 40, 3786-3801.

[27] Padilha Jr, E.; R. de Pelegrini Soares, Cardozo, N. Analysis of equations of state for polymer, *Polímero,s* **2015,** *25*-33.